\documentclass{article}[11pt]
\usepackage{graphicx}
\usepackage{amssymb}
\usepackage{epstopdf}
\usepackage{amsmath}
\usepackage{natbib}
\usepackage{subfigure}
\usepackage{geometry}
\usepackage{lscape}

\begin{document}

\begin{center}
{\large {Comparing the Nano-Resolution Depth Sensor to the Co-located Ocean Bottom Seismometer at MARS}}\\\vspace{1.6mm}
Elena Tolkova$^1$, Theo Schaad$^2$ \\\vspace{1mm}
{\emph{$^1$NorthWest Research Associates}}\\\vspace{0.5mm}
{\emph{$^2$Paroscientific, Inc., and Quartz Seismic Sensors, Inc.}}\\\vspace{2mm}
\end{center}
\bigskip

{\bf{Abstract}}\\ 
Data from earthquake measurements using a Nano-Resolution Pressure Sensor and a co-located Ocean Bottom Seismometer have been compared and analyzed. It is shown that the Nano-Resolution Pressure Sensor can serve as a vertical OBS. Moreover, using the pressure sensor in conjunction with a seismometer helps to separate the ocean and seismic signals.\\

{\bf{Background}}\\
The key sensing instruments used for detecting tsunamis are inherently digital pressure transducers manufactured by Paroscientific, Inc. Advances in counting circuitry and signal processing have improved the resolution of these quartz crystal resonator pressure sensors and accelerometers to a sensitivity of parts-per-billion (nano-resolution) over an extended spectrum. These sensors are used in remote and cabled tsunami warning applications including the NOAA DART System and the Japan Trench Tsunami Observation System [1].

In 2010-2011, the first field tests of a nano-resolution pressure sensor were conducted at the Monterey Accelerated Research System (MARS) cabled observatory at a depth of about 900 meters. Comparisons were made to the standard resolution pressure sensor used in the DART System's bottom pressure recorder. The nano-resolution depth sensor has orders of magnitude improved resolution and is able to distinguish pressure changes equivalent to a few microns of equivalent water depth that are the result of microseisms and longer-period infra-gravity waves. Previously undetectable micro-tsunamis were measured and pressure signals generated by earthquakes were compared to land-based seismic instruments [2, 3].

This note analyzes and compares the pressure signals from the nano-resolution pressure sensor to the measurements of the vertical velocity of the ocean floor from a co-located high-broadband seismometer (Guralp CMG-40T - Quanterra Q330) tested at the same time by UCSD for the Ocean Observatory Initiative. The two sets of tests provided the unique opportunity to relate the movements of the sea floor and variations of the pressure above the ocean bottom, and identify physical phenomena detected by either sensor.\\

{\bf{Analysis of the Observations}} \\
The pressure depth sensor reported at a 40 sample-per-second rate and had flat response throughout. 
The co-located OBS had flat response to velocity from 0.1 Hz (10 s) to 50 Hz. The instrument reported at 200 sps (sample-per-second).
Two seismic events have been recorded by these stations:
\begin{itemize}
\item
October 21, 2010 17:53 M6.9 event in the Gulf of California, 1810 km from the MARS site;
\item
March 11, 2011 05:46 M9.0 event near the east coast of Honshu, Japan, 8010 km from the site.
\end{itemize}
The seismic and pressure records of the above events were processed to isolate the same frequency band.
The seismic signals (records of the vertical velocity of the sea floor) were band-pass filtered and processed for instrument response removal with the SAC package from IRIS in a band from 0.016 Hz (60 s period) to 20 Hz. The pressure signals were de-tided by curve-fitting and then high-pass filtered with Butterworth filter with a cut-off at 60 s period. \\

The most apparent conclusion from visual examination of the records is that the nano-resolution pressure sensor had detected vibrations of the sea floor as clearly as the seismometer did (Fig. 1). 
Different types of waves can be observed in the spectrograms of the records (Fig. 2). The spectrograms were computed as power spectral density of windowed signal fragments, function of frequency and the position of the time window. The 180-s-long raised cosine window was moved through a 6 s interval, and the frequency resolution is about 0.01 Hz.
Higher frequency P-waves arrived about 17:57 on 2010/10/21 and about 05:58 on 2011/03/11 with a central  frequency near 0.3 Hz. The lower frequency S-wave train arrived about 18:02 (2010) and 06:20 (2011). The leading frequency of the S-waves is the lowest at the head of the wave-train and gradually increases toward its tail. This dispersive behavior is clearly seen in the spectrograms of each record. In 2011, the S-waves exhibit two leading frequencies. \\

Pressure variations display all the features of the seismic signal, which indicates that the pressure variations are induced by the vertical sea floor  motion. P-waves seem to be more efficient in transmitting into water, since their amplitudes relative to S-waves are higher in the pressure records. 
Microseisms (0.15-0.3 Hz) are also transmitted from the sea floor to water. 
The pressure variations caused by infra-gravity waves in the ocean (0.02 Hz and lower) induce ground vibrations seen on the seismometer.\\

{\bf{Perspectives for Further Study}} \\
This brief analysis of the field data opens some quantitative questions that should be studied further. What is the transfer function between the measurements of the co-located seismic and pressure sensors for different types of waves, both seismic waves and ocean waves? 
What is the adequate theoretical model for the pressure signal induced by a seismic wave traveling under the ocean floor? How well can 
the seismometer response to infra-gravity water waves be eliminated using the measurements of a co-located pressure sensor?\\

Our first attempts to eliminate the infra-gravity signal in the seismic record are presented in Figure \ref{remov}. The figure shows the spectrograms of the OBS records in the 2010 and 2011 events before and after the infra-gravity signal had been removed. The removal operation uses the data from the co-located pressure sensor and an in-house data processing algorithm (under development).
In the smaller local 2010 event, S-waves penetrated the infra-gravity band for a short time (around 18:00 on 2010/10/21), while in the extreme 2011 event, S-waves occupied the infra-gravity band for many hours. The removal results can be seen when not obscured by S-waves.
Overall, 95 \% of the seismic record variance induced by the ocean infra-gravity waves was removed with the help of a pressure record, while the seismic signal in the same frequency band remained intact. \\
\newpage 
{\bf{Summary}}\\
The analysis clearly shows that the nano-resolution depth sensor is a good vertical ocean-bottom seismometer in the absence of other instrumentation. In conjunction with a seismometer it can be used to separate the seismic motion at longer periods from infra-gravity waves, subject to further study. The depth sensor alone provides measurements of long-period processes, such at tsunamis. In combination, co-located seismic and pressure sensors in the vicinity of the tsunami originating area provide sufficient data for early detection, evaluation and forecasting earthquake triggered tsunamis. \\

{\emph{Co-located nano-resolution pressure sensors, accelerometers, and tiltmeters provide the best geodetic measurements for the analysis and prediction of geophysical disasters including distant and near-field tsunamis.}}

\bigskip
{\bf{References}}
\begin{enumerate}
\item
J. Paros, P. Migliacio, T. Schaad, Nano-Resolution Sensors for Disaster Warning Systems, IEEE Conference Publishing, OCEANS 2012, Yeosu, Korea, May 2012.
\item
J. Paros, E. Bernard, J. Delaney, C. Meinig, M. Spillane, P. Migliacio, L. Tang, W. Chadwick, T. Schaad, and S. Stalin, Breakthrough underwater technology holds promise for improved local tsunami warnings, IEEE Conference Publishing, UT11+SSC11, Tokyo, Japan, March 2011.
\item
J. Paros, P. Migliacio, T. Schaad, C. Meinig, M. Spillane, L. Tang, S. Stalin, and W. Chadwick, Nano-resolution technology demonstrates promise for improved tsunami warnings on the MARS project, IEEE Conference Publishing, OCEANS 2012, Yeosu, Korea, May 2012.
\end{enumerate}

\begin{landscape}
\begin{figure}[ht]
	\resizebox{1.4\textwidth}{!}
		{\includegraphics{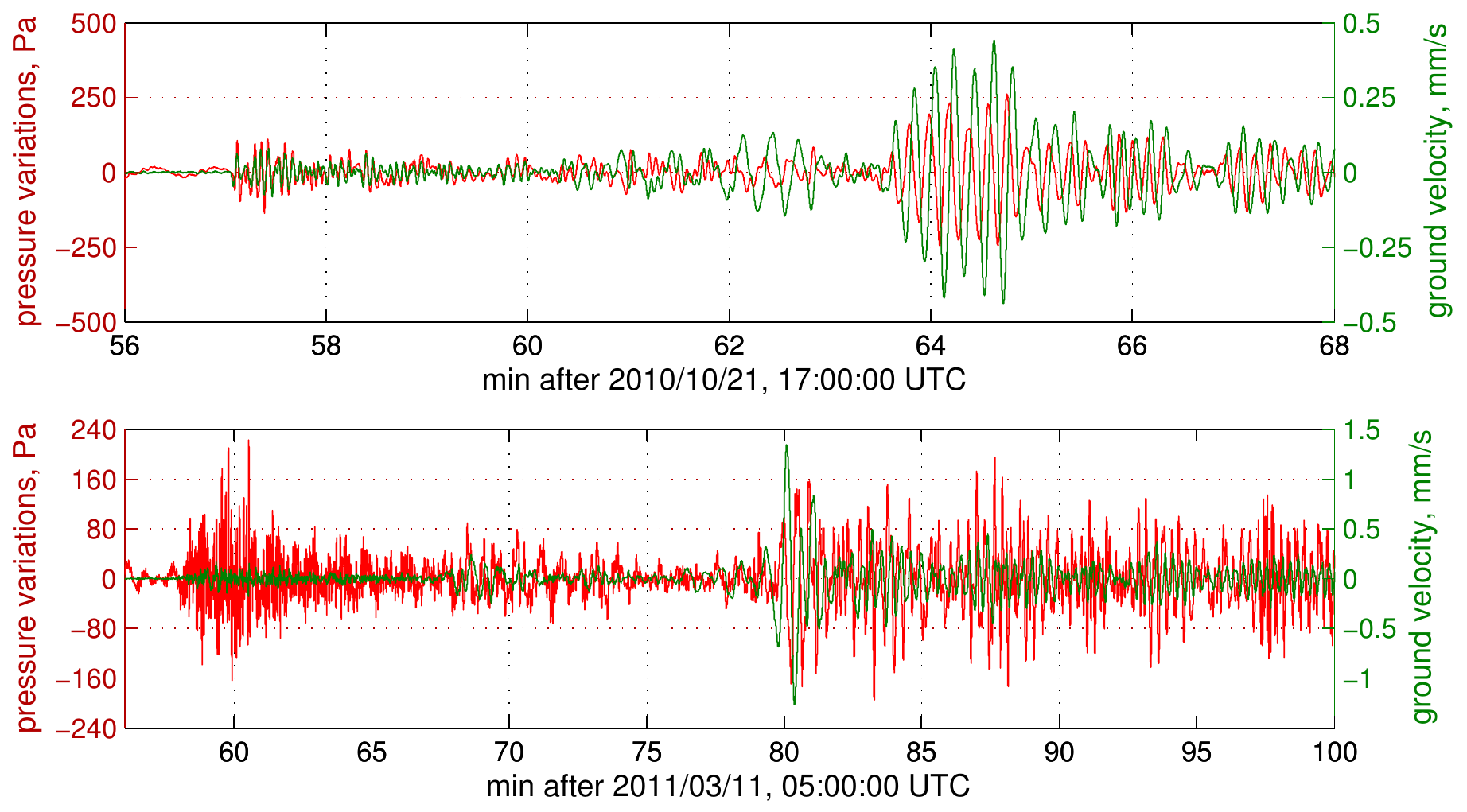}}
	\caption{Records of the vertical component of the ground velocity and of the bottom pressure variations (HP filtered) upon seismic waves arrival at MARS after October 21, 2010, and March 11, 2011 earthquakes. }
	\label{records}
\end{figure} 
\end{landscape}

\begin{figure}[ht]
\begin{tabular}{c}
	\includegraphics[height=0.6\textwidth]{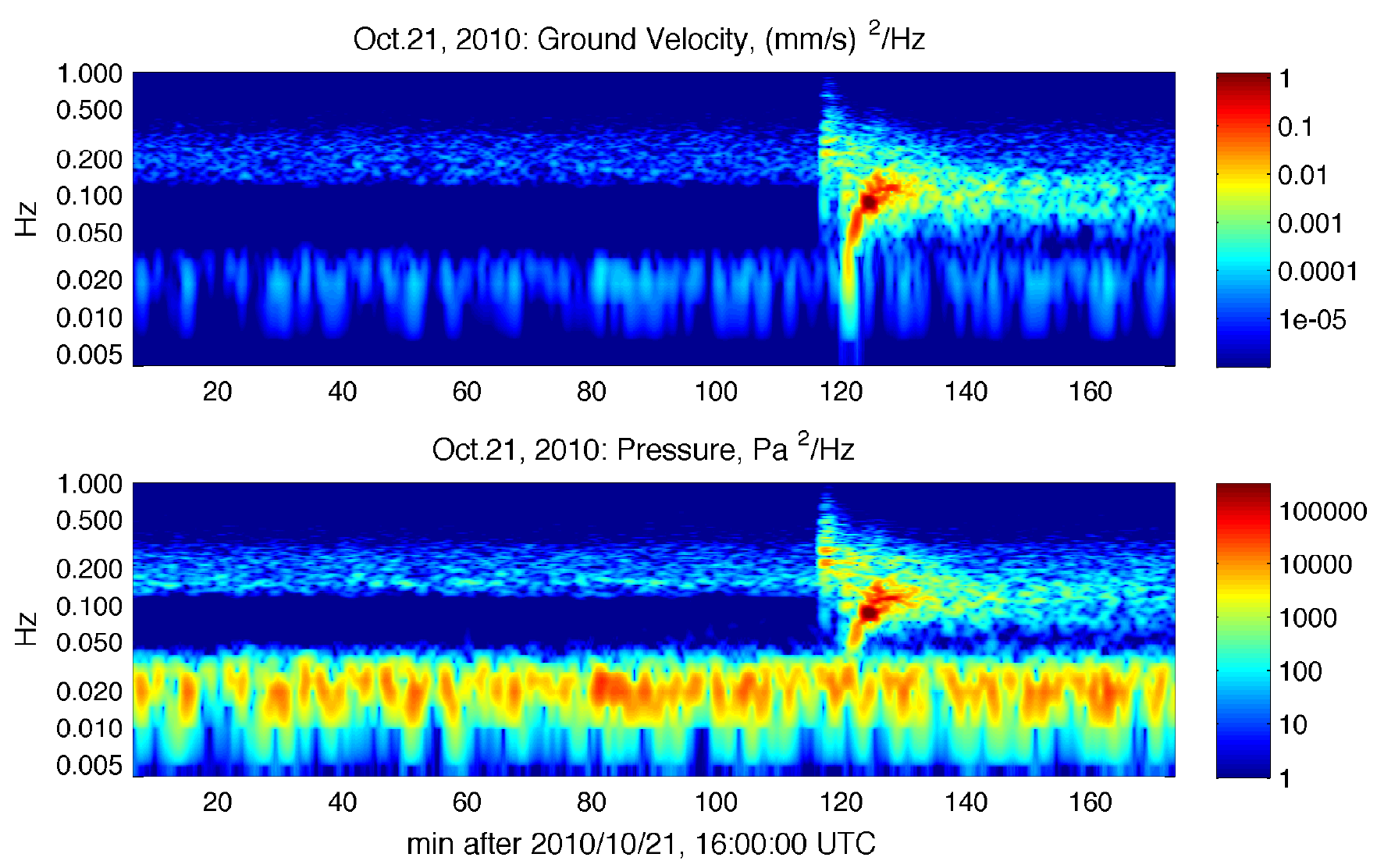} \\
	\includegraphics[height=0.6\textwidth]{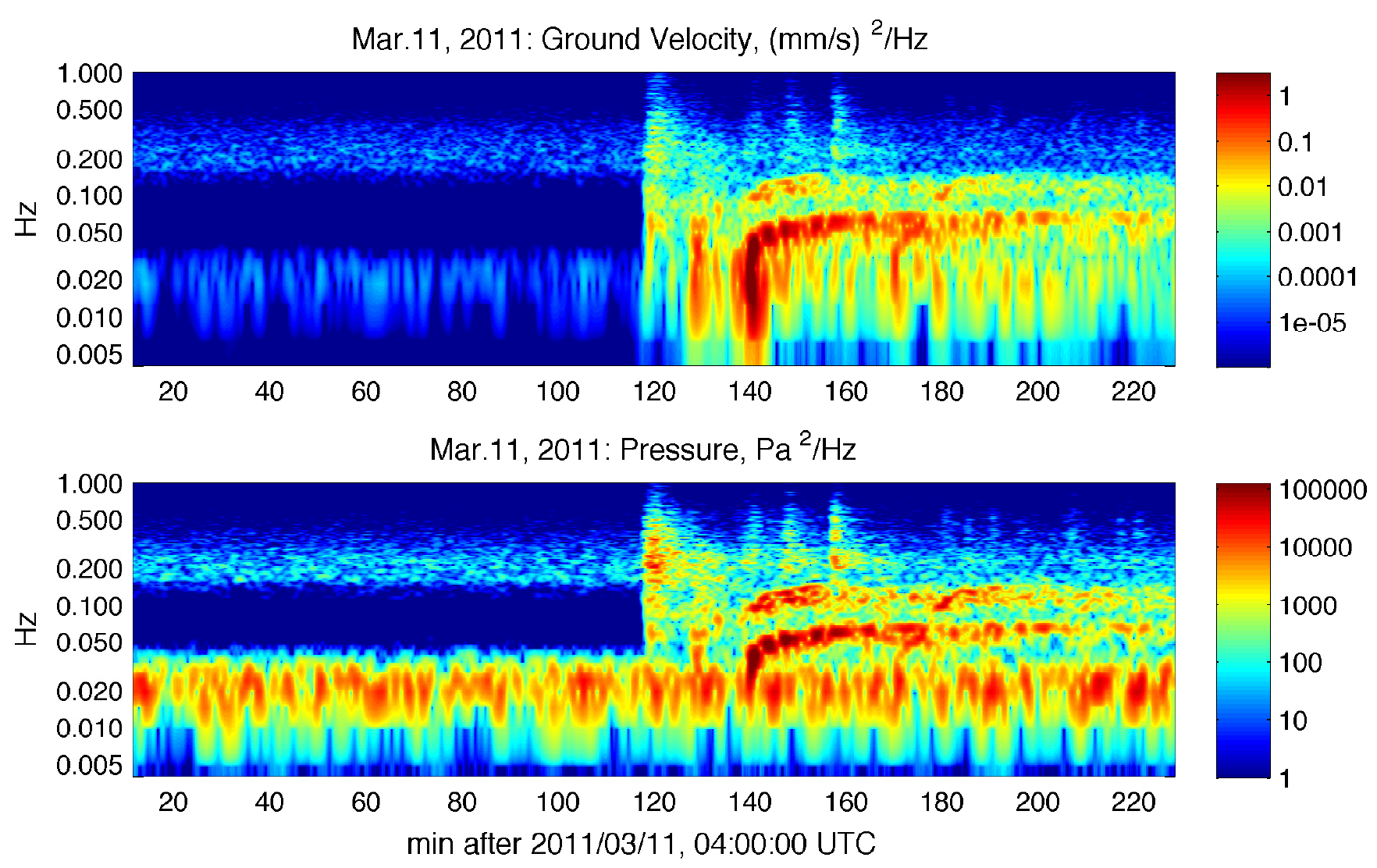}
\end{tabular}
	\caption{The spectrograms of the ocean floor velocity (top) and the pressure variations (bottom) in the 2010 (top) and 2011 (bottom) events.}
	\label{spectra}
\end{figure} 

\begin{figure}[ht]
\begin{tabular}{c}
	\includegraphics[height=0.6\textwidth]{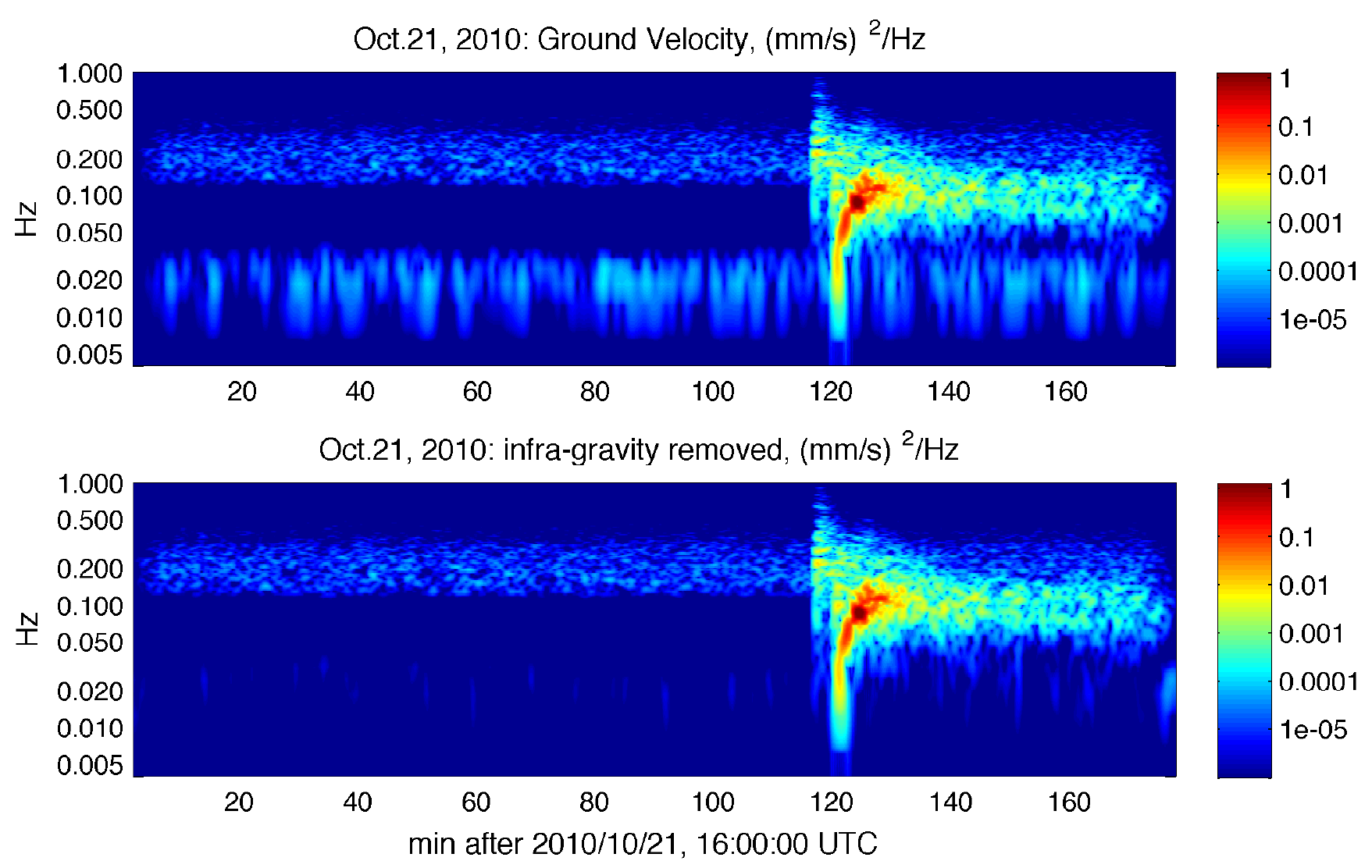} \\
	\includegraphics[height=0.6\textwidth]{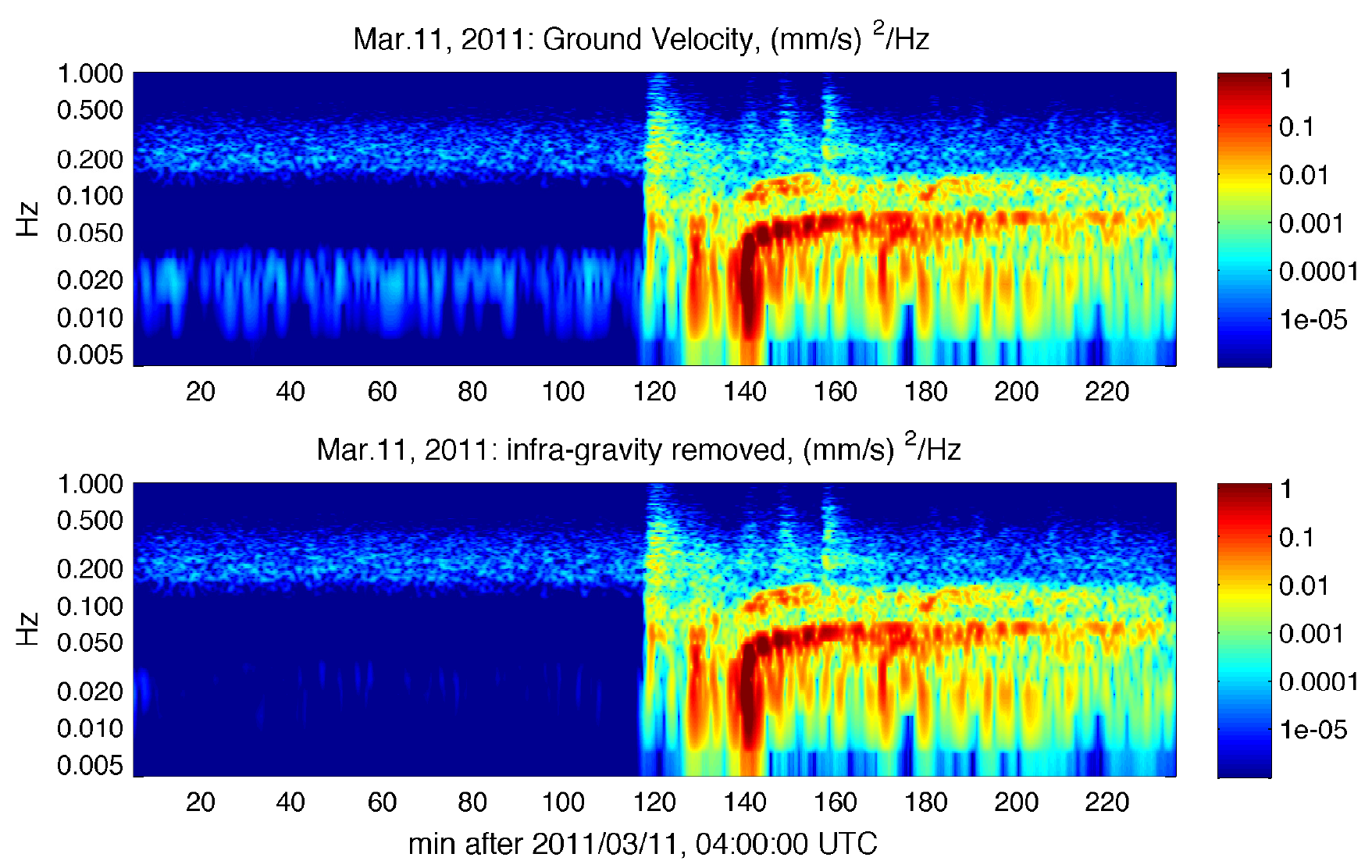}
\end{tabular}
	\caption{The spectrogram of the OBS record before (upper pane in each pair) and after (lower pane in each pair) the infra-gravity signal had been removed using data from the co-located pressure sensor, in the 2010 (top) and 2011 (bottom) events.}
	\label{remov}
\end{figure} 

\end{document}